\documentclass[a4paper]{jpconf}
\usepackage{graphicx}
\begin{document}
\title{Relativistic Viscous Hydrodynamics for Multi-Component Systems with Multiple Conserved Currents}

\author{Akihiko Monnai, Tetsufumi Hirano}

\address{Department of Physics, The University of Tokyo, Tokyo 113-0033, Japan}

\ead{monnai@nt.phys.s.u-tokyo.ac.jp}

\begin{abstract}
We would like to formulate relativistic dissipative hydrodynamics for multi-component systems with multiple conserved currents. This is important for analyses of the hot matter created in relativistic heavy ion collisions because particle creations and annihilations of various particle species are frequently taking place there. We show that consistent formulation in such systems involves many non-trivialities, and derive constitutive equations that satisfy Onsager reciprocal relations and describe the systems without ambiguity.
\end{abstract}

\section{Introduction}
One of the most important discoveries of Relativistic Heavy Ion Collider experiments is that relativistic ideal hydrodynamic models quantitatively describe the collective behavior of the quark-gluon plasma (QGP) created in the collisions \cite{Hirano:2008hy}. The success is a great motivation to take non-equilibrium processes into account to explain small discrepancies between theoretical predictions and experimental data. 
Large Hadron Collider experiments also expect heavy ion collisions of higher energies, and viscous hydrodynamic models would become more important as one might observe \textit{less} strongly-coupled QGP because of asymptotic freedom in QCD. 

The ambiguities of viscous hydrodynamic models lie in the formalism of relativistic dissipative hydrodynamics itself. If one considers only the first order dissipative correction to the entropy current, theories are known to become acausal and unstable \cite{HiscockLindblom}. Thus varieties of theories with the second order correction have been proposed. However, the resulting constitutive equations for the dissipative currents differ among theories. Also, most viscous hydrodynamic models are based on the theories for single component systems with binary collisions. Considering that the QGP is a multi-component system with particle creation and annihilation, it is important that one establishes a multi-component formalism which takes account of inelastic scatterings. This naturally motivates us to consider multiple conserved currents because one needs to solve continuity equations for conserved charges, rather than the ones for particle numbers.

\section{Formulation of Relativistic Dissipative Hydrodynamics}
We would like to formulate relativistic dissipative hydrodynamics for multi-component and multi-conserved current systems \cite{Monnai:2010qp}. Relativistic hydrodynamics is a general theoretical framework which describes collective motion of strongly-coupled systems with the energy-momentum conservation $\partial_\mu T^{\mu \nu} = 0$, the charge conservations $\partial_\mu N_J^\mu = 0$ and the law of increasing entropy $\partial_\mu s^\mu \geq 0$. Here the index $J$ denotes different types of conserved quantum number. Typically, one can consider the baryon number $B$ or the strangeness $S$ for such quantities in relativistic heavy ion collisions. 

We introduce tensor decompositions of the energy-momentum tensor $T^{\mu \nu}$ and the conserved currents $N_J^\mu$ in terms of the flow field $u^\mu$ to define thermodynamic quantities
\begin{eqnarray}
\label{eq:em_eqnoneq}
T ^{\mu \nu} &=& (e_0+\delta e) u^\mu u^\nu - (P_0+\Pi ) \Delta ^{\mu \nu} + W^{\mu} u^{\nu} + W^{\nu} u^{\mu} + \pi^{\mu \nu} , \\
\label{eq:n_eqnoneq}
N_J ^{\mu} &=& (n_{J0}+ \delta n_J) u^{\mu} + V_J^{\mu},
\end{eqnarray}
where $\Delta^{\mu \nu} = g^{\mu \nu} - u^\mu u^\nu$ is the projection operator. There are $2+N$ equilibrium quantities: energy density $e_0$, hydrostatic pressure $P_0$ and type-$J$ charge density $n_{J0}$. On the other hand, one has $10+4N$ dissipative currents: bulk pressure $\Pi$, energy density distortion $\delta e$, energy current $W^\mu$, shear stress tensor $\pi ^{\mu \nu}$, type-$J$ charge density distortion $\delta n_{J}$ and type-$J$ charge current $V_J^\mu$. Although thermodynamic stability conditions \cite{Monnai:2009ad, Israel:1979wp} require $\delta e$ and $\delta n_{J}$ to be zero, we have to keep them for the moment to correctly count the number of the dissipative currents.
In ideal hydrodynamics, the conservation laws and the equation of state $P_0 = P_0 (e_0, \{ n_{J0} \})$ give sufficient number of equations to describe systems because the number of unknowns is $5+N$ in such systems. 
In dissipative hydrodynamics, one needs \textit{constitutive equations} because there are the $10+4N$ additional unknowns $\Pi$, $\delta e$, $W^\mu$, $\pi^{\mu \nu}$, $\delta n_J$ and $V_J^\mu$. We would like to derive them from the law of increasing entropy, because it embodies irreversible processes. 

The energy-momentum tensor and the conserved currents are expressed in relativistic kinetic theory with the phase space distribution $f^i$ as
\begin{eqnarray}
\label{eq:emcc}
T^{\mu \nu} &=& \sum _i \int \frac{g_i d^3 p}{(2\pi )^3 E_i} p_i^\mu p_i^\nu f^i , \ N_J^{\mu} = \sum _i \int \frac{q_i^J g_i d^3 p}{(2\pi )^3 E_i} p_i^\mu f^i ,
\end{eqnarray}
where $g_i$ and $q_i^J$ are degeneracy and type-$J$ quantum number of $i$-th particle species, respectively. 
We introduce the higher moment equations for the energy-momentum conservation and for the charge conservations 
\begin{eqnarray}
\label{eq:ene_consv_high}
\partial _\alpha I^{\mu \nu \alpha} &=& \sum _ i \int \frac{g_i d^3 p}{(2\pi )^3 E_i} p_i^\mu p_i^\nu p_i^\alpha \partial _\alpha f^i = Y^{\mu \nu}, \\
\label{eq:chg_consv_high}
\partial _\alpha I^{\mu \alpha}_J 
&=& \sum _ i \int \frac{q_i^J g_i d^3 p}
{(2\pi )^3 E_i} p_i^\mu p_i^\alpha \partial _\alpha f^i = Y_J^{\mu},
\end{eqnarray}
to derive second order constitutive equations. $Y^{\mu \nu}$ and $Y_J^{\mu}$ are to be determined from the law of increasing entropy. The conventional Israel-Stewart theory \cite{Israel:1979wp} has only Eq.~(\ref{eq:ene_consv_high}) because single component systems with binary collisions are considered. However, one has to introduce the new sets of equations (\ref{eq:chg_consv_high}) to consistently match the numbers of equations with that of dissipative currents in multi-component and/or multiple conserved current systems. 

The entropy production can be written in kinetic theory as
\begin{eqnarray}
\label{eq:ent_pro}
\partial _\mu s^\mu = - \partial _\mu \sum_i \int \frac{g_i d^3 p}{(2\pi )^3 E_i}
p_i ^\mu \bigg[ f^i\ln f^i - \frac{1}{\epsilon}
(1 + \epsilon f^i) \ln (1 + \epsilon f^i) \bigg] = \sum _i \int \frac{g_i d^3 p_i}{(2\pi )^3 E_i} p_i^\mu y^i \partial_\mu f^i ,
\end{eqnarray}
where $y^i$ is defined in $f^i = [\exp (y^i) - \epsilon]^{-1}$. The sign factor $\epsilon$ is $+1$ for fermions and $-1$ for bosons. We estimate the distortion of distribution $\delta f^i = f^i -f_0^i$ in terms of dissipative currents by generalizing the Grad's moment method \cite{Israel:1979wp} for the systems with multiple conserved currents. If we require that the resulting constitutive equations satisfy Onsager reciprocal relations \cite{Onsager1, Onsager2}, the only possible moment expansion would be 
\begin{eqnarray}
\label{eq:dym}
\delta f^i &=& - f_0^i (1 + \epsilon f_0^i) (p_i^\mu \sum _J q_i^J \varepsilon _{\mu}^J + p_i^\mu p_i^\nu \varepsilon _{\mu \nu}) ,
\end{eqnarray} 
where $\varepsilon _{\mu}^J$ and $\varepsilon _{\mu \nu}$ are self-consistently determined so that the distribution reproduces the energy-momentum tensor and the conserved currents (\ref{eq:emcc}). Then the $\varepsilon$'s are expressed as linear combinations of the dissipative currents. On the other hand, the entropy production (\ref{eq:ent_pro}) is expressed in terms of $\varepsilon$'s and $Y$'s as 
\begin{eqnarray}
\label{eq:ent_pro2}
\partial _\mu s^\mu 
&=& \sum_J \varepsilon^J_\nu Y_J^\nu 
+ \varepsilon_{\nu \rho} Y^{\nu \rho} \geq 0.
\end{eqnarray}
The law of increasing entropy requires that $Y$'s are linear combinations of $\varepsilon$'s. Then the second order constitutive equations for multi-component systems with multiple conserved currents are obtained by writing the moment equations (\ref{eq:ene_consv_high}) and (\ref{eq:chg_consv_high}) with the dissipative current:
\begin{eqnarray}
\Pi &=& -\zeta \nabla _\mu u^\mu - \zeta_{\Pi \delta e} D\frac{1}{T} + \sum _J \zeta_{\Pi \delta n_J} D\frac{\mu_J}{T} \nonumber \\
&-& \tau_\Pi D \Pi + \sum_J \chi_{\Pi \Pi}^{aJ} \Pi D\frac{\mu _J}{T} + \chi_{\Pi \Pi}^b \Pi D \frac{1}{T} + \chi_{\Pi \Pi}^c \Pi \nabla _\mu u^\mu \nonumber \\
&+& \sum_J \chi_{\Pi W}^{aJ} W_\mu \nabla ^\mu \frac{\mu _J}{T} + \chi_{\Pi W}^b W_\mu \nabla ^\mu \frac{1}{T} + \chi_{\Pi W}^c W_\mu D u ^\mu + \chi_{\Pi W}^d \nabla ^\mu W_\mu \nonumber \\
&+& \sum_{J,K} \chi_{\Pi V_J}^{aK} V^J_\mu \nabla ^\mu \frac{\mu_{K}}{T} + \sum_J \chi_{\Pi V_J}^b V^J_\mu \nabla ^\mu \frac{1}{T} + \sum_J \chi_{\Pi V_J}^c V^J_\mu D u ^\mu + \sum_J \chi_{\Pi V_J}^d \nabla ^\mu V^J_\mu \nonumber \\
&+& \chi_{\Pi \pi} \pi _{\mu \nu} \nabla ^{\langle \mu} u^{\nu \rangle} ,
\label{eq:Pi}
\\
W^\mu &=& -\kappa _W \bigg( \frac{1}{T} D u^\mu + \nabla ^\mu \frac{1}{T} \bigg) + \sum_J \kappa_{W V_J} \nabla ^\mu \frac{\mu _J}{T}\nonumber \\
&-& \tau_{W} \Delta^{\mu \nu} D W_\nu + \sum_J \chi_{W W}^{aJ} W^\mu D\frac{\mu _J}{T} + \chi_{WW}^b W^\mu D \frac{1}{T} \nonumber \\
&+& \chi_{W W}^c W^\mu \nabla _\nu u^\nu + \chi_{WW}^d W^\nu \nabla _\nu  u^\mu + \chi_{WW}^e W^\nu \nabla ^\mu u_\nu \nonumber \\
&-& \sum_J \tau _{W V_J} \Delta^{\mu \nu} D V^J_\nu + \sum_{J,K} \chi_{WV_{J}}^{aK} V_J^\mu D\frac{\mu _{K}}{T} + \sum_J \chi_{WV_J}^b V_J^\mu D \frac{1}{T} \nonumber \\
&+& \sum_J \chi_{W V_J}^c V_J^\mu \nabla ^\nu u_\nu + \sum_J \chi_{WV_J}^d V_J^\nu \nabla _\nu  u^\mu + \sum_J \chi_{WV_J}^e V_J^\nu \nabla ^\mu u_\nu \nonumber \\
&+& \sum_J \chi_{W \pi}^{aJ} \pi^{\mu \nu} \nabla_\nu \frac{\mu _J}{T} + \chi_{W \pi}^b \pi^{\mu \nu} \nabla_\nu \frac{1}{T} + \chi_{W \pi}^c \pi ^{\mu \nu} D u_\nu + \chi_{W \pi}^d \Delta^{\mu \nu} \nabla ^\rho \pi _{\nu \rho} \nonumber \\
&+& \sum_J \chi_{W \Pi}^{aJ} \Pi \nabla ^\mu \frac{\mu _J}{T} + \chi_{W \Pi}^b \Pi \nabla ^\mu \frac{1}{T} + \chi_{W \Pi}^c \Pi D u^\mu + \chi_{W \Pi}^d \nabla ^\mu \Pi ,
\label{eq:W}
\\
V_J^\mu &=& \sum_{K} \kappa_{V_J V_K} \nabla ^\mu \frac{\mu _K}{T} - \kappa _{V_J W} \bigg( \frac{1}{T} D u^\mu + \nabla ^\mu \frac{1}{T} \bigg) \nonumber \\
&-& \sum_{K} \tau _{V_J V_{K}} \Delta^{\mu \nu} D V^{K}_\nu + \sum_{K,L} \chi_{V_J V_{K}}^{aL} V_{K}^\mu D \frac{\mu _{L}}{T} + \sum_{K} \chi_{V_J V_{K}}^b V_{K}^\mu D \frac{1}{T} \nonumber \\
&+& \sum_{K} \chi_{V_J V_{K}}^c V_{K}^\mu \nabla _\nu u^\nu + \sum_{K} \chi_{V_J V_{K}}^d V_{K}^\nu \nabla _\nu u^\mu + \sum_{K} \chi_{V_J V_{K}}^e V_{K}^\nu \nabla ^\mu u_\nu \nonumber \\
&-& \tau_{V_J W} \Delta^{\mu \nu} D W_\nu + \sum_{K} \chi_{V_J W}^a W^\mu D \frac{\mu _{K}}{T} + \chi_{V_J W}^b W^\mu D \frac{1}{T} \nonumber \\
&+& \chi_{V_J W}^c W^\mu \nabla ^\nu u_\nu + \chi_{V_J W}^d W^\nu \nabla _\nu u^\mu + \chi_{V_J W}^e W^\nu \nabla ^\mu u_\nu \nonumber \\
&+& \sum_{K} \chi_{V_J \pi}^{aK} \pi^{\mu \nu} \nabla_\nu \frac{\mu _{K}}{T} + \chi_{V_J \pi}^b \pi^{\mu \nu} \nabla_\nu \frac{1}{T} + \chi_{V_J \pi}^c \pi ^{\mu \nu} D u_\nu \nonumber + \chi_{V_J \pi}^d \Delta^{\mu \nu} \nabla ^\rho \pi _{\nu \rho} \\
&+& \sum_{K} \chi_{V_J \Pi}^{aK} \Pi \nabla ^\mu \frac{\mu _{K}}{T} + \chi_{V_J \Pi}^b \Pi \nabla ^\mu \frac{1}{T} + \chi_{V_J \Pi}^c \Pi D u^\mu + \chi_{V_J \Pi}^d \nabla ^\mu \Pi ,
\label{eq:V}
\\
\pi^{\mu \nu} &=& 2 \eta \nabla ^{\langle \mu} u^{\nu \rangle} \nonumber \\
&-& \tau_\pi D \pi^{\langle \mu \nu \rangle} + \sum_J \chi_{\pi \pi}^{aJ} \pi^{\mu \nu} D \frac{\mu_J}{T} + \chi_{\pi \pi}^b \pi^{\mu \nu} D \frac{1}{T} + \chi_{\pi \pi}^c \pi ^{\mu \nu} \nabla _\rho u^\rho + \chi_{\pi \pi}^d \pi ^{\rho \langle \mu} \nabla _\rho u^{\nu \rangle} \nonumber \\
&+& \sum_J \chi_{\pi W}^{aJ} W^{\langle \mu} \nabla ^{\nu \rangle} \frac{\mu_J}{T} + \chi_{\pi W}^b W^{\langle \mu} \nabla ^{\nu \rangle} \frac{1}{T} + \chi_{\pi W}^c W^{\langle \mu} D u ^{\nu \rangle} + \chi_{\pi W}^d \nabla ^{\langle \mu} W^{\nu \rangle} \nonumber \\
&+& \sum_{J,K} \chi_{\pi V_J}^{aJ} V_J^{\langle \mu} \nabla ^{\nu \rangle} \frac{\mu_{K}}{T} + \sum_J \chi_{\pi V_J}^b V_J^{\langle \mu} \nabla ^{\nu \rangle} \frac{1}{T} + \sum_J \chi_{\pi V_J}^c V_J^{\langle \mu} D u^{\nu \rangle} + \sum_J \chi_{\pi V_J}^d \nabla ^{\langle \mu} V_J ^{\nu \rangle} \nonumber \\
&+& \chi _{\pi \Pi} \Pi \nabla ^{\langle \mu} u^{\nu \rangle} .
\label{eq:pi}
\end{eqnarray}
Here $D=u^\mu \partial_\mu$ and $\nabla^\mu = \Delta^{\mu \nu} \partial_\nu$ are time- and space-like derivatives. The angle brackets denote the operation $A^{\langle \mu} B^{\nu \rangle}
= [ \frac{1}{2} (\Delta ^\mu _{\ \alpha} \Delta ^\nu _{\ \beta} + \Delta ^\mu _{\ \beta} \Delta ^\nu _{\ \alpha}) - \frac{1}{3}\Delta^{\mu \nu} \Delta_{\alpha \beta}] A^{\alpha} B^{\beta}$. $\zeta$'s, $\kappa$'s and $\eta$ are first order transport coefficients; $\zeta$ is bulk viscosity, $\kappa_{WW}$ energy conductivity, $\kappa_{V_J V_J}$ charge conductivity and $\eta$ shear viscosity. $\tau$'s are relaxation times, and $\chi$'s are second order transport coefficients. The ratios of the first and the second order coefficients can be determined in kinetic theory.

\section{Discussion and Conclusions}

We have derived constitutive equations for multi-component systems with multiple conserved currents in arbitrary frame. We would like to emphasize four points. Firstly, the equations have Onsager cross terms, although most of the conventional formalisms do not consider them carefully. These terms contain important physics such as Soret and Dufour effects, which are chemical diffusion caused by thermal gradient and its counter-effect respectively. The reciprocal relations are satisfied. 
Secondly, there are relaxation terms -- terms with time-like derivatives of the dissipative currents themselves -- 
in the constitutive equations. These terms play important roles in preserving causality because they cause exponential damping of the dissipative currents. 
Thirdly, we have independent equations for energy and charge currents. This means that our formalism can describe systems in an arbitrary frame unlike the conventional formalisms in which one has to choose either energy or particle frames.
Fourthly, we have several second order terms which do not appear in Ref.~\cite{Israel:1979wp}. 
For further detailed discussion including comparison with other approaches \cite{Baier:2007ix, Tsumura:2009vm, Betz:2009zz}, see Ref.~\cite{Monnai:2010qp}.

\ack 
The authors acknowledge fruitful discussions with T.~Hatsuda, T.~Kodama, T.~Koide, T.~Kunihiro and S.~Muroya.
The work of A.M. is supported by JSPS Research Fellowships for Young Scientists.
The work of T.H. was partly supported by Grant-in-Aid for Scientific Research No.~19740130 and by Sumitomo Foundation No.~080734.

\end{document}